\documentclass[a4paper,11pt]{article}
\pdfoutput=1 

\usepackage{jcappub} 
               
\usepackage{breakurl}        
\usepackage{longtable}  

\usepackage[T1]{fontenc} 

\title{\boldmath Relative contribution of the magnetic field barrier and solar wind speed in ICME-associated Forbush decreases}

\author[a,1]{Ankush Bhaskar,\note{Corresponding author.}}
\author[b,2]{Prasad Subramanian,\note{also at Center for Excellence in Space Sciences, India, http://www.cessi.in}}
\author[a]{Geeta Vichare}

\affiliation[a]{Indian Institute of Geomagnetism,\\ Kalamboli Highway, New Panvel, Navi Mumbai-
410218, India.}
\affiliation[b]{Indian Institute of Science Education and Research, Dr. Homi Bhabha Road, Pashan, Pune 411008, India}

\emailAdd{ankushbhaskar@gmail.com}
\emailAdd{p.subramanian@iiserpune.ac.in }
\emailAdd{vicharegeeta@gmail.com}

\abstract{
We study 50 cosmic ray Forbush decreases (FDs) from the Oulu neutron monitor data during 1997-2005 that were associated with Earth-directed interplanetary coronal mass ejections (ICMEs). Such events are generally thought to arise due to the shielding of cosmic rays by a propagating diffusive barrier. The main processes at work are the diffusion of cosmic rays across the large-scale magnetic fields carried by the ICME and their advection by the solar wind. In an attempt to better understand the relative importance of these effects, we analyse the relationship between the FD profiles and those of the interplanetary magnetic field (B) and the solar wind speed (Vsw). Over the entire duration of a given FD, we find that the FD profile is generally well (anti)correlated with the B and Vsw profiles. This trend holds separately for the FD main and recovery phases too. For the recovery phases, however, the FD profile is highly anti-correlated with the Vsw profile, but not with the B profile.  While the total duration of the FD profile is similar to that of the Vsw profile, it is significantly longer than that of the B profile. 
}

\begin{document}
\maketitle
\flushbottom

\section{Introduction}
\label{sec:intro}
Long term and transient modulations of galactic cosmic rays due to solar activity range from long-term solar cycle-associated modulations \cite{nagashima1980long,usoskin1998correlative,cane1999cosmic,aslam2012solar} to transient modulations called Forbush decreases (FDs) \citep{Forbush1937}. Our focus in this paper will be on FDs, which are generally thought to arise due to the shielding of cosmic rays by a propagating turbulent magnetic barrier \citep{lockwood1986characteristic,badruddin2002transient,candia2004,subramanian2009,Babu_et_al_2013,raghav2014understanding}.
The magnetic barrier can be an Interplanetary Coronal Mass Ejection (ICME) and its associated shock or the interaction region between the slow and fast solar wind, known as Co-rotating Interaction Region(CIR). 
Despite numerous attempts to relate properties of FDs with measured parameters of ICMEs/CIRs at 1 AU \cite{richardson1996relationship,belov2001determines,belov2008forbush,Richardson&cane2011,dumbovic2011cosmic,dumbovic2012cosmic,belov2014coronal} there are significant gaps in our understanding of their underlying physical mechanisms.

Our broad understanding of cosmic ray propagation is based on Parker's transport equation \cite{parker1965passage}. 
Interplanetary magnetic fields can often be compressed and distorted by shock waves from coronal mass ejections (CMEs); this causes a shielding effect against incident cosmic rays. Therefore, diffusion and gradient drifts are considered to be important contributors to the FD phenomenon \cite{parker1965passage,le1991simulation}.
The importance of the turbulent sheath region ahead of the CME in determining the FD magnitude has been emphasised previously (e.g.\cite{giacalone1999transport,subramanian2009,Babu_et_al_2013,raghav2014understanding}). Many models neglect the convective term \cite{forman1971convection} and consider a diffusion-only scenario for high energy cosmic rays. However, the contribution of solar wind convection needs to be accounted to explain the observed FD profiles. In this regard various observational studies have pointed to the role of the solar wind convection in modulation of the cosmic rays (e.g. \cite{iucci1979high,richardson1996relationship,aslam2012solar}). Nonetheless, the relative contributions of the diffusion and advection terms is still a matter of debate.
ICMEs expand rapidly as they propagate \cite{michalek2009expansion,subramanian2014self}, and this expansion can advect cosmic ray particles.

 The main goal of the current study is to understand the relative importance of diffusion and solar wind convection in causing the FD phenomenon. In order to do so, we use data from the Oulu neutron monitor and study the correlations between FD profiles and those of the (corresponding) interplanetary magnetic field (B) and the solar wind speed ($V_{sw}$).


The rest of the paper is arranged as follows: section 2 describes the data, observations and analysis methodology. We present some representative results from applying a specific diffusion-convection model to one event in section 3.  Section 4 ends the paper with discussions and conclusions.

\section{Database, methods and results}
Our starting point is the list of interplanetary coronal mass ejections (ICMEs) compiled by Richardson and Cane \cite{richardson2010near} for 1997--2005\footnote{also available at http://www.srl.caltech.edu/ACE/ASC/DATA/level3/icmetable2.htm}. 
We then look for FDs corresponding to these ICMEs using neutron monitor data from the  Oulu neutron monitor (Geographic coordinates: 65.05$^o$ N, 25.47$^o $  E, vertical geomagnetic cut-off rigidity 0.8 GV).
\subsection{Event selection}
We only consider events that exhibit the well-acknowledged characteristics of FDs - a transient decrease followed by a gradual recovery. Very complex FDs or events contaminated with Ground level Enhancements (GLEs) were discarded. Specifically, we shortlist events whose time profiles show some correspondence with the B and Vsw time profiles. 
Interplanetary parameters such as the solar wind speed and magnetic field were obtained from in-situ observations at 1 AU available at CDAWEB\footnote{http://cdaweb.gsfc.nasa.gov/}. The Vsw, B and neutron flux data were smoothened to 60 min time resolution using the moving average method. We shortlist 50 events using these criteria. Details of these events are presented in Table \ref{table:tab_one}. The information includes the onset and end of each ICME (from \cite{richardson2010near}) as well as the onset and end of the associated FD. 

\subsection{Representative example event}

\begin{figure}[h]
\begin{center}
\includegraphics[width = 18 cm]{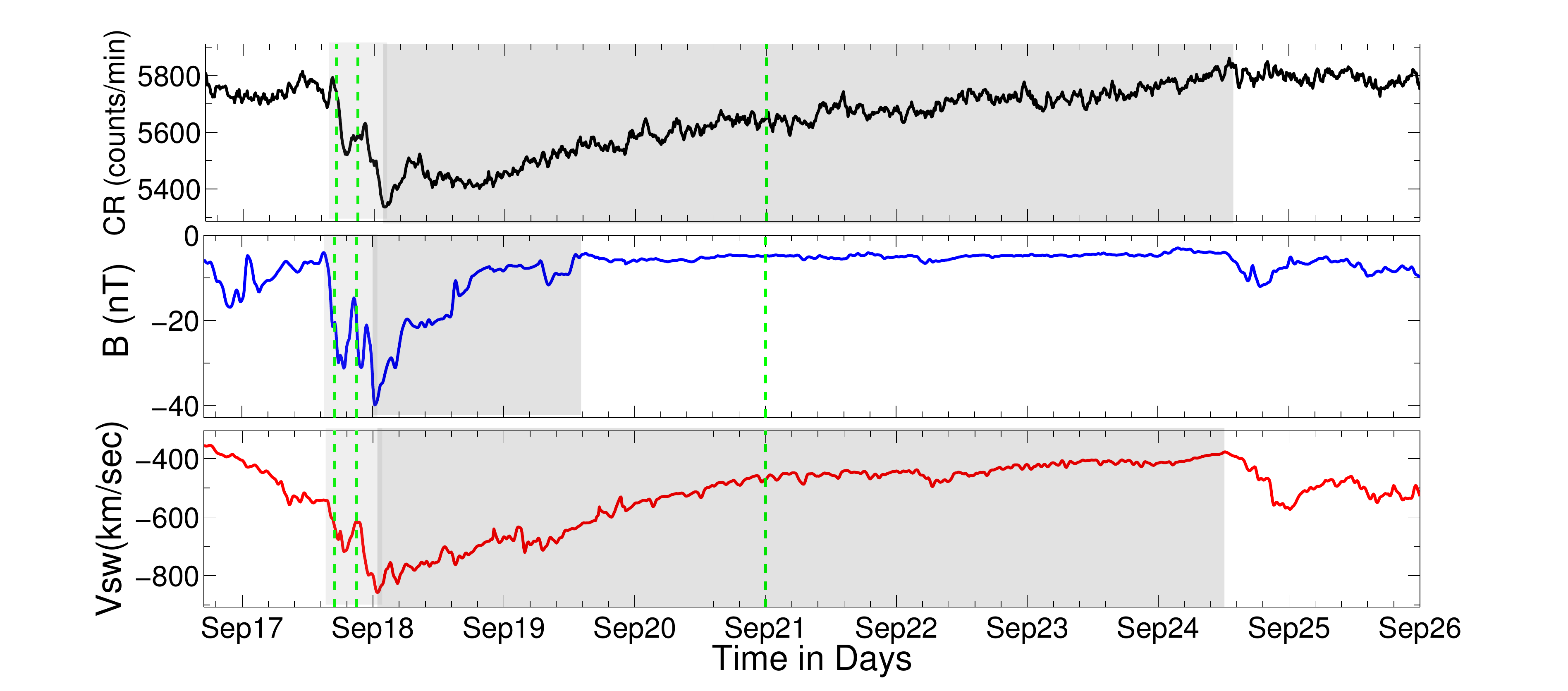}
\caption{ Forbush decrease event of September 17, 2000. The light gray/dark gray shaded region indicates the main/recovery phase seen in respective quantities. The vertical dashed lines from left to right are representative of onset of disturbance, start of ICME and end of ICME respectively. Note that the profiles of B and Vsw are inverted.}
\label{fig:event_time}
\end{center}
\end{figure}

Before presenting results for all the events in our database, we illustrate our methodology using a representative example is the FD event of September 17, 2000. The neutron flux (FD), Vsw and B profiles for this event are shown in Figure \ref{fig:event_time}.
In order to facilitate comparison with the FD profile, the Vsw and B profiles are inverted. The initial increase in the solar wind speed (Vsw) (depicted as a decrease in Figure~\ref{fig:event_time}) is a signature of the arrival of the ICME and its associated shock. The interplanetary magnetic field (B) is frozen into the solar wind plasma; the density compression in the shock-sheath region therefore increases the magnetic field, which is depicted as a decrease in Figure~\ref{fig:event_time}. The gradual recovery of Vsw is due to the expansion of the ICME flux rope \cite{russell2003icme,russell2005defining}. Due to expansion of ICME, the leading edge of the ICME travels faster than rear, which shows up as the gradual decrease in Vsw. The recovery in B, on the other hand, is much faster than that of Vsw, and is presumably related to the decrease in the density subsequent to the shock-sheath region.
The vertical lines denote the arrival of interplanetary shock, the ICME start time and the ICME end respectively. By analogy with geomagnetic storm temporal profiles, we divide the FD profile into two parts: the main phase and recovery phase \citep{gonzalez1994geomagnetic,kudela2004cosmic}. The main phase is characterized by a sharp decrease (lasting a few hours) in the cosmic ray flux whereas the recovery phase is characterized by a gradual recovery (lasting for a few days) to the pre-onset flux. The main phase of the FD is defined as the part of the profile between the quiescent pre-event level and the first minimum.  The recovery phase is the part of the profile between the first minimum and the point where the profile has (largely) recovered to its pre-event value. The main phase of the FD is shown in light gray and the recovery phase is colored in dark gray. The main and recovery phases for Vsw and B are defined separately, using the same logic. For instance, the main phase for Vsw is the part between the pre-event value of Vsw and its first minimum, while its recovery phase is the part between the first minimum of Vsw and the point where it has attained its pre-event value. Visual inspection of the three panels in Figure ~\ref{fig:event_time} suggests that the main phases of the FD, B and Vsw profiles are generally similar. However, the time taken by B to recover to its pre-event value is the smallest; this is followed by the recovery time for Vsw, and the recovery time for the FD is the longest. These trends are generally representative for the rest of the events, and are quantified as explained below.


\subsection{Results: similarity between FD, Vsw and B profiles}
The similarity between the FD, B and Vsw profiles is quantified using a time-lag correlation of the form

\begin{equation}
 R = \frac{n\sum xy-(\sum x)(\sum y)} { \sqrt{ n (\sum x^{2})- (\sum x)^{2} } \sqrt{ n (\sum y^{2})- (\sum y)^{2} }  }
\end{equation}

\begin{figure}[h]
\begin{center}
\includegraphics[width = 17 cm]{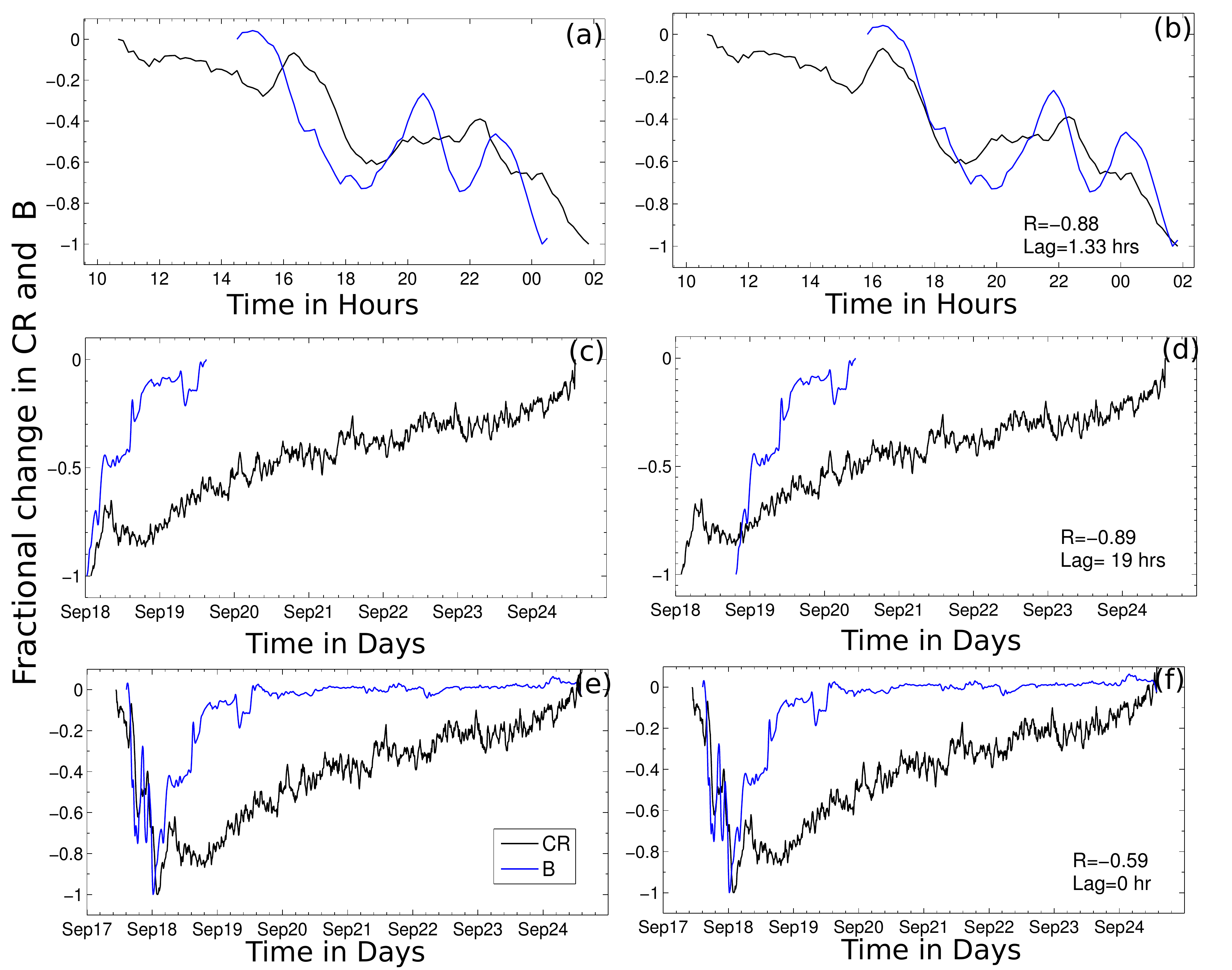}
\caption{Correlated profiles of B (blue) and neutron flux (black) from top for FD even of September 17, 2000: main phase, recovery phase and complete profile respectively. The left (a,c,e) panels  show original profiles whereas, the right (b,d,f) panels show profiles which are time shifted by maximum lag. The maximum cross-correlation coefficient (R) and associated lag (in hours) is indicated in each right panel.}
\label{fig:corr_B}
\end{center}
\end{figure}

\begin{figure}[h!]
\begin{center}
\includegraphics[width = 17 cm]{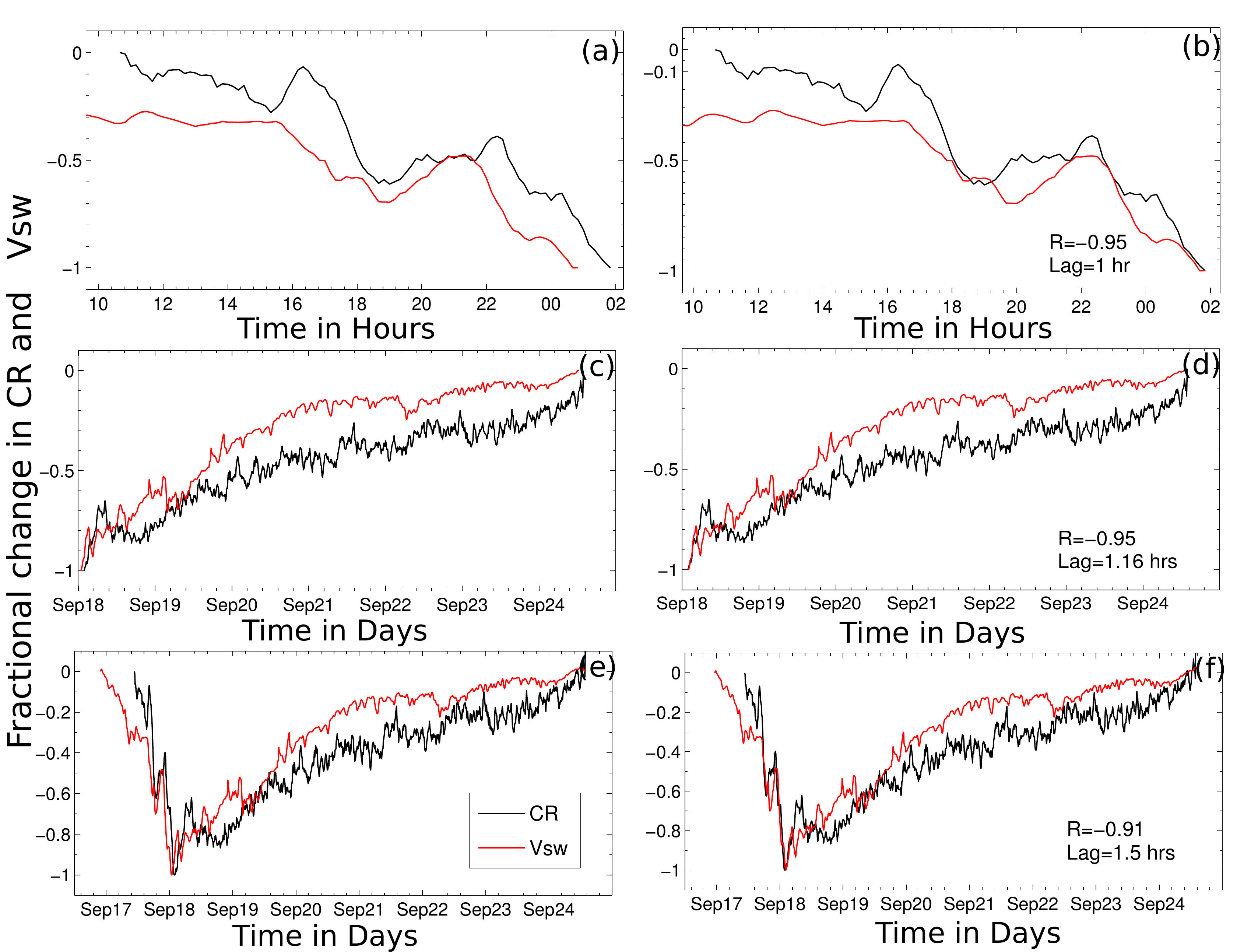}
\caption{Correlated profiles of Vsw (red) and neutron flux (black) from top for FD even of September 17, 2000: main phase, recovery phase and complete profile respectively. The left (a,c,e) panels show original profiles whereas, the right (b,d,f) panels show profiles which are time shifted by maximum lag. The maximum cross-correlation coefficient (R) and associated lag (in hours) is indicated in each right panel.}
\label{fig:corr_Vsw}
\end{center}
\end{figure}

We carry out three kinds of cross-correlations: 
\begin{enumerate}
\item
FD with Vsw and FD with B over the complete duration of the FD
\item
Main phase of the FD with the main phase of Vsw and also with the main phase of B
\item
Recovery phase of the FD with the recovery phase of Vsw and with the recovery phase of B
\end{enumerate}
In each case, the correlation coefficient attains a maximum value for a particular value of the time lag. The maximum correlation coefficient and the corresponding time lag are noted.
 Examples for the event of figure~\ref{fig:event_time} are shown in figures~\ref{fig:corr_B} and \ref{fig:corr_Vsw}. For each quantity (e.g., neutron flux), the profile represents the percentage deviation from a suitable pre-event baseline. Figure ~\ref{fig:corr_B} shows the correlation between the FD and B profiles for a) the entire duration of the FD event, b) the main phase of the FD and the main phase of B and c) the recovery phase of the FD and the recovery phase of B. The FD and Vsw profiles are similarly compared in Figure~\ref{fig:corr_Vsw}.

\begin{figure}[h!]
\begin{center}
\includegraphics[width = 17 cm]{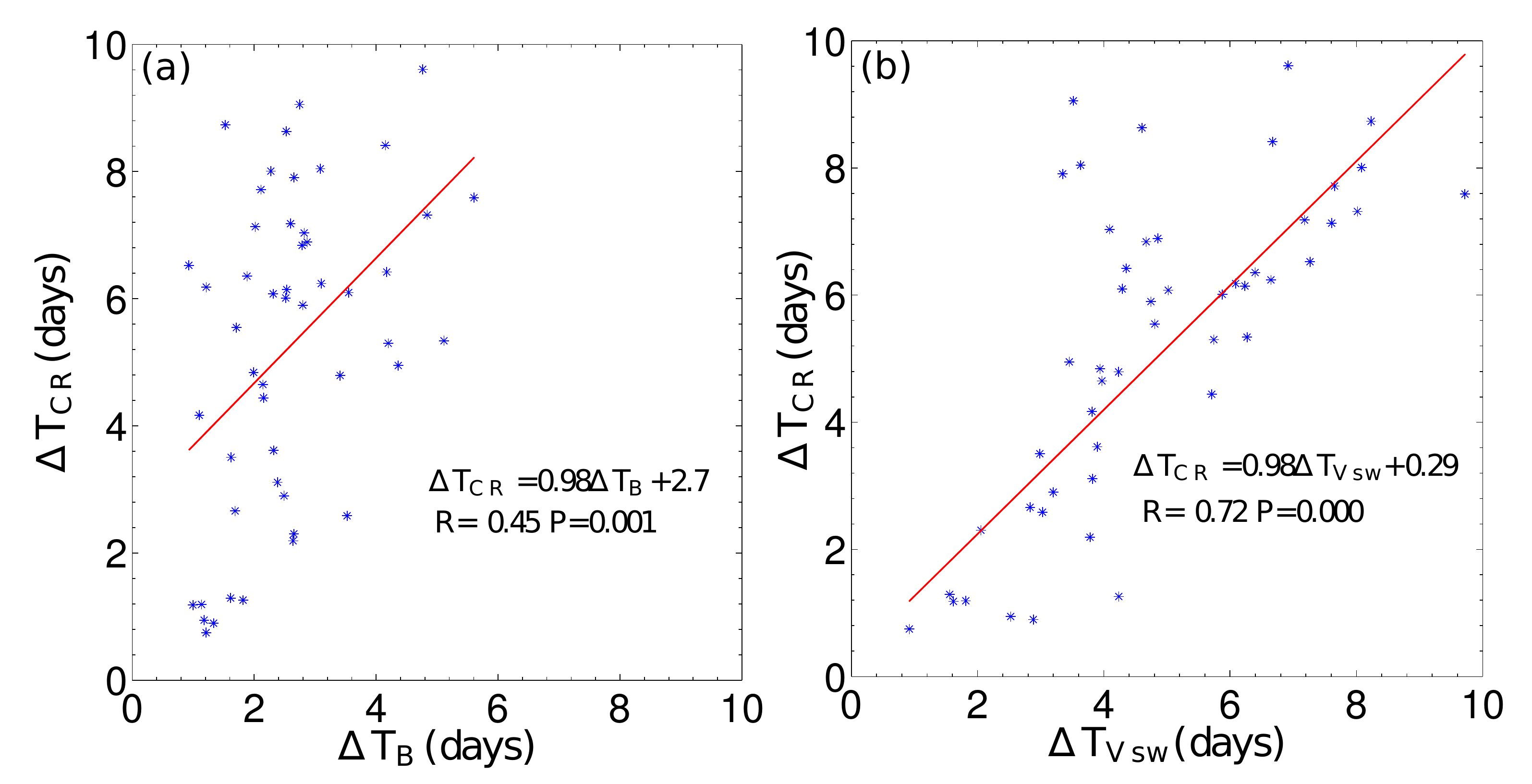}
\caption{Relatinship between total duration of Forbush decrease observed in neutron flux ($\Delta T_{CR}$) and corresponding interplanetary disturbance observed in interplanetary magnetic field ($\Delta T_{B}$) and solar wind speed ($\Delta T_{Vsw}$). (a) The total duration in B is poorly correlated with that observed in the neutron flux. (b) The total disturbance duration observed in Vsw is well correlated with total duration of Forbush decrease in neutron flux. The maximum cross-correlation coefficient (R) and associated significance level (P value) is indicated in the each panel.}
\label{fig:totaldur}
\end{center}
\end{figure}

We carry out this exercise for each of the shortlisted events listed in table~\ref{table:tab_one}. In each case, the maximum correlation coefficient and the corresponding time lag is summarized in table~\ref{table:tab_two}. 
The general trends revealed by this exercise are as follows: i) Over the entire duration of the FD, the neutron flux is well (anti) correlated with B and Vsw. ii) This statement is generally true for the main phase too iii) During the recovery phase, however, the anti-correlation between the neutron flux and Vsw is noticeably more pronounced than that between the neutron flux and B. As noted briefly earlier, the duration of the recovery phase for B is shorter than for the neutron flux and Vsw. Figure \ref{fig:totaldur} shows scatter-plots between a) the (neutron flux) FD duration ($\Delta T_{CR}$) and the duration of the corresponding B profile ($\Delta T_{B}$) and b) the (neutron flux) FD duration ($\Delta T_{CR}$) and the duration of the corresponding Vsw profile ($\Delta T_{Vsw}$). The durations are measured for the entire event, not only for the main or recovery phases. We note that $\Delta T_{CR}$ and $\Delta T_{Vsw}$ are well correlated (correlation coefficient = 0.72); furthermore, since the intercept of the fit between these two quantities is close to unity, $\Delta T_{CR} \approx \Delta T_{Vsw}$. Conversely, $\Delta T_{CR}$ is poorly correlated with $\Delta T_{B}$. This is primarily due to the fact that the recovery phase for B is much shorter than that for the neutron flux.

\section{Diffusion-convection model}
 Before we conclude, we briefly point out the application of the well established diffusion-convection model to one of our events. \cite{richardson1996relationship} showed that enhanced solar wind convection is a possible cause for FDs associated with high-speed solar wind streams. Differentiating Eq 2 of their paper yields the following equation that relates the depression in particle counts to variations in the solar wind speed:

\begin{equation}
 \frac{\partial u}{u} = -3CN \frac{\partial V_{sw}}{c}
\label{eqdc}
\end{equation}

In equation~\eqref{eqdc} u is cosmic ray density, C is the Compton-Getting factor, $V_{sw}$ is solar wind bulk speed, N is a number of mean free paths of diffusing cosmic ray protons. Further details of the model can be found in \cite{richardson1996relationship}. As an illustrative example, we apply this model to the event shown in Figure~\ref{fig:event_time}.

\begin{figure}[h!]
\begin{center}
\includegraphics[width = 14 cm]{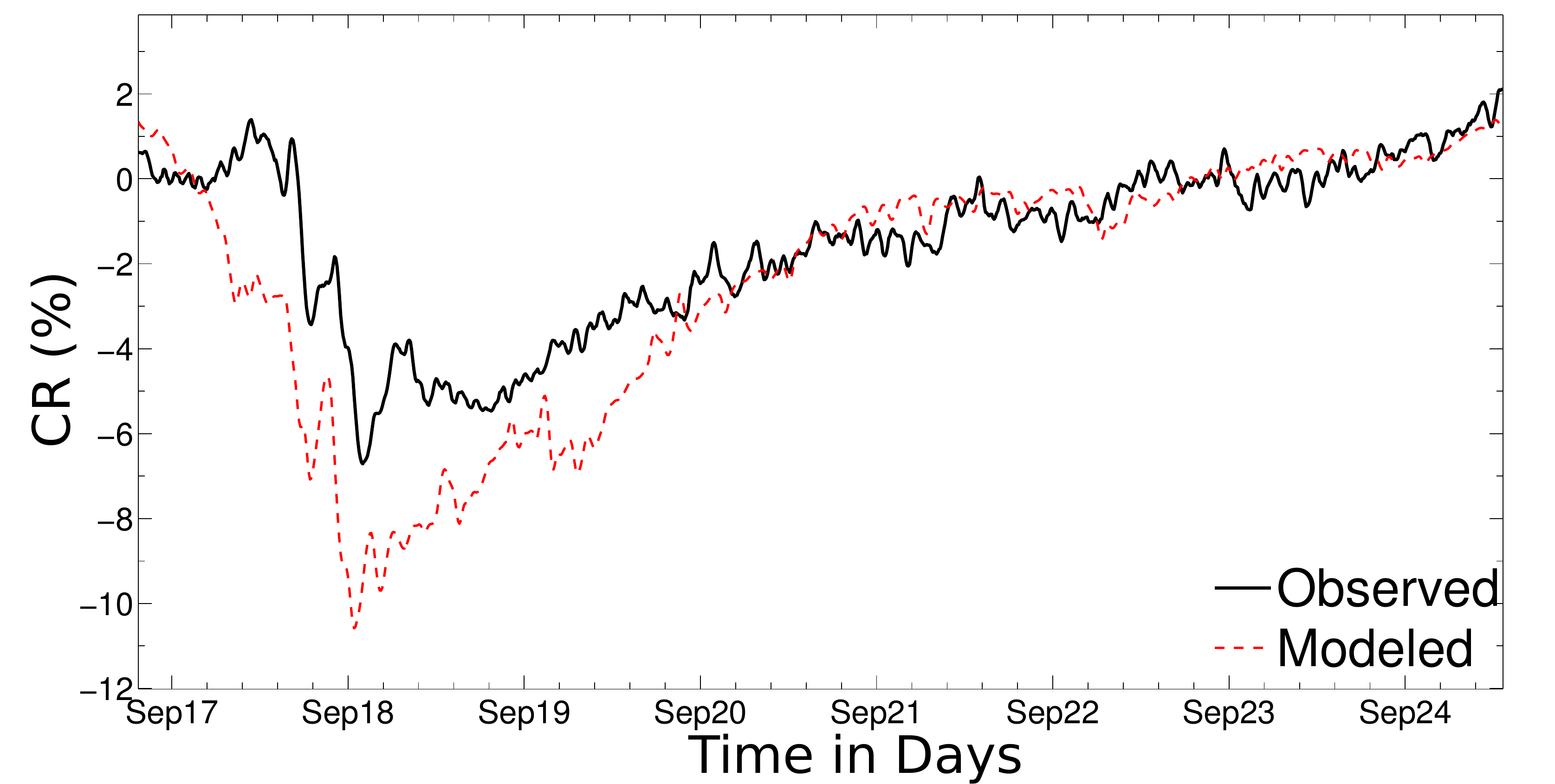}
\caption{The observed (black line) and modeled (red dashed line) profiles of FD took place on September 17, 2000. The convection-diffusion model is fitted to the recovery phase of FD and mean free path (N=14) was estimated for minimum $\chi^2$ (see the text for more details).}
\label{fig:model}
\end{center}
\end{figure}

Since the Oulu neutron monitor has cutoff rigidity of 0.8 GV and an effective energy of 5.6 GeV \cite{alanko2003effective}, we have adopted C=1.8 \cite{gleeson1968compton}. The constant N essentially corresponds to the number of mean free paths. We treat N as a free parameter which is constrained as follows: we varied N from 1 to 25, and fitted the model to the observed recovery profile of the FD event. We fit the recovery profile, since this is the part that seems to anti-correlate best with $V_{sw}$. The best fit is obtained by minimizing $\chi^2 = \sum \limits_{i}  \frac{(E_{i}-O_{i})^2}{Var_{i}} $ value, where $E_{i}$ is the estimated CR variation and $O_{i}$  is the observed CR variation. This procedure yields N=14. Figure~\ref{fig:model} shows the model for the complete FD (main phase + recovery phase).  It is evident from figure ~\ref {fig:model} that the model mimics the FD profile in the recovery phase well, when the interplanetary magnetic field has settled to its pre-event value. This implies that the magnetic field might have less to do the modulation of CR during the recovery phase as compared to the solar wind speed. It may be noted that the model overestimates the  cosmic ray flux depression in the main phase. The overestimation by the Convection-diffusion model during the main phase is intriguing. N can be thought of as distance(r)/mean free path($\lambda$); if one assumes $\lambda \sim r^{0.4}$, then $N \sim r^{0.6}$ implying that N increases with radial distance even though $\lambda$ increases with radial distance as well. During the main phase of FD, the value of N is smaller since the ICME is smaller. As the ICME expands beyond 1 AU (when the recovery phase of FD is observed), the ICME increases in size, leading to an increased N for the recovery phase. Thus model estimates using a higher value of N (estimated using the recovery phase) can overestimate CR variations during the main phase. To know whether model fitted value of N is realistic, one can estimate the possible value of mean free path of cosmic rays ($\lambda$). \cite{burlaga1983dynamical}  have shown that individual stream structure diminishes considerably by $\sim 8.5$ AU. We assume that interplanetary structure until $\sim 10$ AU can affect the variations in cosmic rays observed at the Earth and that $\lambda$ is independent of heliocentric distance. Therefore, estimated $\lambda$ is $\sim 0.6$ AU. However, in reality $\lambda$ is dependent on distance i.e proportional to $r^{0.4}$, then $\lambda$ (1 AU) turns out to be $\sim 0.25$ AU at 1 AU \cite{hamilton1977radial}. This value for $\sim 6$ GeV protons is consistent with the observed $\lambda\sim 0.30$ at radial distance of 1 AU  by \cite{bieber1994proton}. Interestingly, these estimates of $\lambda$ are similar to those reported by {\cite{richardson1996relationship}} for FDs associated with CIRs implying the importance of solar wind speed for both kinds of FDs.	

\section{Discussion and Conclusions}
\normalsize

FDs occur due to (partial) shielding of galactic cosmic rays by the propagating, diffusive barrier arising from Earth-directed ICMEs. This involves the competing processes of cosmic ray diffusion across the barrier formed by the magnetic field enhancement of the ICME-shock sheath structure and advection by the solar wind. The increase in B (which inhibits cross-field cosmic ray transport) is accompanied by an increase in the solar wind speed (which advects cosmic rays and thus aids transport). Thereafter, the weakening of the magnetic field (which aids cross-field cosmic ray transport) is accompanied by a decrease in the solar wind speed (which decreases advection and inhibits cosmic ray transport).

We have shortlisted a set of 50 carefully selected FDs from the Oulu neutron monitor data that were associated with Earth-directed ICMEs. For each of these FDs, we study the relation between the FD profile and the solar wind speed (Vsw) and interplanetary magnetic field (B) profiles associated with the ICME. For each profile, the part between the pre-event background and the first minimum is called the main phase, and the part between the minimum and the post-event background is called the recovery phase. For all the events, we find that the main phase of the FD profiles are anti-correlated with the main phases of Vsw and B to a reasonably good extent. On the other hand, the recovery phase of the FD profiles show marked anti-correlation only with the recovery phase of the Vsw profiles, and not with those of the B profiles. We attempted to explain the observed FD profile for one representative event using the diffusion-convection model. The model explains the observed profile of recovery phase for N=14 implying $\lambda$ (1 AU) $\sim 0.25$ AU which is consistent with the past reports by \cite{bieber1994proton,richardson1996relationship}. Furthermore, the total durations of the FD events (measured from pre-event background to post-event background) are similar to the total durations of the corresponding Vsw profile. The B profile durations, on the other hand, are considerably shorter, and show no significant correlation with the FD durations. The competing processes of cosmic ray diffusion across magnetic fields and advection by the solar wind are at work in both the main and recovery phases of the FD. The detailed analysis in this paper is expected to contribute towards understanding the relative contributions of these processes.

\acknowledgments

The solar wind parameters, interplanetary magnetic field and geomagnetic indices used in this paper are obtained from  
CDAWEB. We thank ACE science center, Oulu neutron monitor for making data available in public domain. 



\clearpage
%

\clearpage
\footnotesize
\begin{longtable}{|lllll|lll|llll|llll|}
\caption[Tab1]{\normalfont \textbf {Observed parameters of the ICMEs and associated \textit{FD} Events}} \label{Tab1} \\
\hline
&\multicolumn{4}{c|}{
\textbf{ICME start}}&\multicolumn{3}{c|}{\textbf{ICME end}}&\multicolumn{4}{c|}{\textbf{FD onset}}&\multicolumn{4}{c|}{\textbf{FD end}} \\
\hline

N0. & YYYY & MM & DD &  HH & MM & DD & HH &  MM & DD &  HH & Min &  MM & DD &  HH & Min \\ 

\hline\hline
1  & 1997 & 4  & 11 & 6  & 4  & 11 & 19 & 4  & 10 & 16 & 19 & 4  & 15 & 2  & 53 \\
2  & 1997 & 8  & 3  & 13 & 8  & 4  & 3  & 8  & 3  & 7  & 32 & 8  & 4  & 13 & 45 \\
3  & 1997 & 10 & 10 & 11 & 10 & 10 & 22 & 10 & 10 & 2  & 13 & 10 & 16 & 12 & 17 \\
4  & 1997 & 11 & 22 & 19 & 11 & 23 & 14 & 11 & 22 & 5  & 54 & 11 & 29 & 2  & 3  \\
5  & 1997 & 12 & 10 & 18 & 12 & 12 & 0  & 12 & 9  & 15 & 5  & 12 & 14 & 10 & 5  \\
6  & 1997 & 12 & 30 & 10 & 12 & 31 & 11 & 12 & 29 & 15 & 49 & 1  & 6  & 8  & 59 \\
7  & 1998 & 3  & 25 & 13 & 3  & 26 & 10 & 3  & 25 & 4  & 28 & 4  & 1  & 8  & 47 \\
8  & 1998 & 5  & 2  & 5  & 5  & 4  & 2  & 5  & 1  & 21 & 33 & 5  & 4  & 4  & 48 \\
9  & 1998 & 5  & 4  & 10 & 5  & 7  & 23 & 5  & 4  & 5  & 15 & 5  & 11 & 19 & 22 \\
10 & 1998 & 8  & 26 & 22 & 8  & 28 & 0  & 8  & 26 & 1  & 3  & 9  & 2  & 8  & 38 \\
11 & 1998 & 9  & 25 & 6  & 9  & 26 & 16 & 9  & 24 & 11 & 57 & 9  & 30 & 9  & 30 \\
12 & 1998 & 10 & 23 & 15 & 10 & 24 & 16 & 10 & 23 & 13 & 3  & 10 & 28 & 9  & 13 \\
13 & 1999 & 1  & 13 & 15 & 1  & 13 & 23 & 1  & 12 & 15 & 30 & 1  & 19 & 12 & 50 \\
14 & 1999 & 2  & 18 & 10 & 2  & 20 & 17 & 2  & 17 & 7  & 49 & 2  & 26 & 9  & 8  \\
15 & 1999 & 6  & 27 & 22 & 6  & 29 & 4  & 6  & 26 & 5  & 32 & 7  & 4  & 20 & 41 \\
16 & 2000 & 3  & 19 & 2  & 3  & 19 & 12 & 3  & 18 & 10 & 48 & 3  & 19 & 15 & 23 \\
17 & 2000 & 4  & 7  & 6  & 4  & 8  & 6  & 4  & 6  & 16 & 29 & 4  & 12 & 20 & 50 \\
18 & 2000 & 5  & 24 & 12 & 5  & 27 & 10 & 5  & 23 & 10 & 9  & 5  & 29 & 10 & 26 \\
19 & 2000 & 6  & 8  & 12 & 6  & 10 & 17 & 6  & 8  & 9  & 13 & 6  & 16 & 10 & 16 \\
20 & 2000 & 6  & 26 & 10 & 6  & 27 & 0  & 6  & 25 & 20 & 6  & 7  & 1  & 9  & 13 \\
21 & 2000 & 9  & 17 & 21 & 9  & 21 & 0  & 9  & 17 & 10 & 35 & 9  & 24 & 13 & 46 \\
22 & 2000 & 12 & 23 & 0  & 12 & 23 & 12 & 12 & 22 & 21 & 1  & 12 & 23 & 14 & 58 \\
23 & 2001 & 3  & 27 & 20 & 3  & 28 & 17 & 3  & 27 & 9  & 10 & 3  & 30 & 11 & 53 \\
24 & 2001 & 3  & 28 & 17 & 3  & 30 & 18 & 3  & 30 & 17 & 38 & 4  & 4  & 9  & 15 \\
25 & 2001 & 3  & 31 & 5  & 3  & 31 & 22 & 3  & 30 & 23 & 40 & 4  & 3  & 14 & 25 \\
26 & 2001 & 4  & 4  & 18 & 4  & 5  & 12 & 4  & 4  & 14 & 29 & 4  & 7  & 12 & 6  \\
27 & 2001 & 4  & 8  & 14 & 4  & 9  & 4  & 4  & 8  & 12 & 38 & 4  & 11 & 4  & 32 \\
28 & 2001 & 4  & 28 & 14 & 5  & 1  & 2  & 4  & 28 & 2  & 32 & 5  & 6  & 20 & 8  \\
29 & 2001 & 5  & 28 & 3  & 5  & 31 & 14 & 5  & 27 & 16 & 37 & 6  & 1  & 15 & 24 \\
30 & 2001 & 8  & 17 & 20 & 8  & 19 & 16 & 8  & 17 & 10 & 8  & 8  & 20 & 22 & 17 \\
31 & 2001 & 10 & 12 & 4  & 10 & 12 & 9  & 10 & 11 & 12 & 38 & 10 & 17 & 16 & 2  \\
32 & 2001 & 10 & 29 & 22 & 10 & 31 & 13 & 10 & 27 & 22 & 24 & 11 & 4  & 20 & 8  \\
33 & 2001 & 11 & 19 & 22 & 11 & 21 & 13 & 11 & 19 & 15 & 21 & 11 & 20 & 14 & 2  \\
34 & 2001 & 11 & 24 & 14 & 11 & 25 & 20 & 11 & 24 & 5  & 30 & 11 & 30 & 14 & 0  \\
35 & 2002 & 3  & 21 & 14 & 3  & 22 & 6  & 3  & 20 & 8  & 5  & 3  & 21 & 5  & 38 \\
36 & 2002 & 4  & 20 & 0  & 4  & 21 & 18 & 4  & 19 & 7  & 52 & 4  & 21 & 12 & 29 \\
37 & 2002 & 9  & 8  & 4  & 9  & 8  & 20 & 9  & 7  & 14 & 41 & 9  & 10 & 4  & 45 \\
38 & 2002 & 10 & 3  & 1  & 10 & 4  & 18 & 10 & 2  & 14 & 27 & 10 & 8  & 16 & 18 \\
39 & 2002 & 12 & 21 & 3  & 12 & 22 & 19 & 12 & 22 & 15 & 13 & 12 & 28 & 17 & 32 \\
40 & 2003 & 3  & 20 & 12 & 3  & 20 & 22 & 3  & 20 & 3  & 40 & 3  & 25 & 11 & 47 \\
41 & 2003 & 10 & 24 & 21 & 10 & 25 & 12 & 10 & 24 & 10 & 48 & 10 & 25 & 15 & 11 \\
42 & 2003 & 11 & 20 & 10 & 11 & 21 & 8  & 11 & 20 & 7  & 44 & 11 & 21 & 14 & 46 \\
43 & 2004 & 7  & 27 & 2  & 7  & 27 & 22 & 7  & 26 & 22 & 14 & 7  & 31 & 2  & 16 \\
44 & 2004 & 11 & 7  & 22 & 11 & 9  & 10 & 11 & 7  & 10 & 22 & 11 & 15 & 10 & 30 \\
45 & 2004 & 11 & 9  & 20 & 11 & 11 & 23 & 11 & 9  & 4  & 41 & 11 & 17 & 14 & 35 \\
46 & 2005 & 1  & 21 & 19 & 1  & 22 & 17 & 1  & 21 & 16 & 31 & 1  & 26 & 23 & 42 \\
47 & 2005 & 5  & 15 & 6  & 5  & 19 & 0  & 5  & 15 & 1  & 10 & 5  & 24 & 15 & 44 \\
48 & 2005 & 5  & 30 & 1  & 5  & 30 & 23 & 5  & 29 & 9  & 35 & 6  & 5  & 10 & 25 \\
49 & 2005 & 8  & 24 & 14 & 8  & 24 & 23 & 8  & 24 & 7  & 29 & 8  & 30 & 13 & 12 \\
50 & 2005 & 9  & 15 & 6  & 9  & 16 & 18 & 9  & 15 & 4  & 4  & 9  & 21 & 16 & 39 \\
\hline
\label{table:tab_one}
\end{longtable}

\scriptsize

\begin{longtable}{|llll|llll|llll|llll|}

\caption[Tab1] {\normalfont \textbf {Cross-correlation of neutron flux with Vsw and  B during total, main and recovery phases of FD	}} \label{Tab2} \\
\hline
&\multicolumn{3}{c|}{\textbf{Event}}&\multicolumn{4}{c|}{\textbf{Total}}&\multicolumn{4}{c|}{\textbf{Main phase}}&\multicolumn{4}{c|}{\textbf{Recovery phase}} \\

\hline
N0. & Y & M & D &  R1-B & Lag &  R1-Vsw & Lag &  R2-B & Lag  & R2-Vsw & Lag &  R3-B & Lag & R3-Vsw &  Lag  \\ 
\hline
&\multicolumn{3}{c|}{\textbf{}}&\multicolumn{1}{c}{\textbf{}}&\multicolumn{1}{c}{\textbf{hrs}}&\multicolumn{1}{c}{\textbf{}}&\multicolumn{1}{c|}{\textbf{hrs}} &\multicolumn{1}{c}{\textbf{}}&\multicolumn{1}{c}{\textbf{hrs}}&\multicolumn{1}{c}{\textbf{}}&\multicolumn{1}{c|}{\textbf{hrs}}& \multicolumn{1}{c}{\textbf{}}&\multicolumn{1}{c}{\textbf{hrs}} &\multicolumn{1}{c}{\textbf{}}&\multicolumn{1}{c|}{\textbf{hrs}} \\

\hline\hline

1  & 1997 & 4  & 10 & \textbf{-0.48} & 6.67  & \textbf{-0.60} & 0.17   & \textbf{-0.92} & 6.83  & \textbf{-0.99} & 2.00   & \textbf{-0.78} & 7.17  & \textbf{-0.77} & -9.00  \\
2  & 1997 & 8  & 3  & \textbf{-0.68} & 0.00  & \textbf{-0.75} & 9.33   & \textbf{-0.79} & -2.33 & \textbf{-0.73} & 9.33   & \textbf{-0.88} & -2.67 & \textbf{-0.88} & 2.00   \\
3  & 1997 & 10 & 10 & \textbf{-0.83} & 0.00  & \textbf{-0.87} & 12.33  & \textbf{-0.69} & 1.83  & \textbf{-0.72} & 12.50  & \textbf{-0.82} & 8.16 & \textbf{-0.86} & 19.50  \\
4  & 1997 & 11 & 22 & \textbf{-0.78} & 8.00  & \textbf{-0.82} & 0.00   & \textbf{-0.89} & 7.83  & \textbf{-0.90} & 7.50   & \textbf{-0.86} & 5.83  & \textbf{-0.85} & -29.33 \\
5  & 1997 & 12 & 9  & \textbf{-0.58} & 4.50  & \textbf{-0.71} & -0.17  & \textbf{-0.60} & 4.83  & \textbf{-0.77} & 4.50   & \textbf{-0.84} & 16.83 & \textbf{-0.77} & 12.33  \\
6  & 1997 & 12 & 29 & \textbf{-0.36} & 0.00  & \textbf{-0.66} & 9.83   & \textbf{-0.67} & -3.83 & \textbf{-0.80} & 7.67   & \textbf{-0.37} & 13.50 & \textbf{-0.61} & 11.50  \\
7  & 1998 & 3  & 25 & \textbf{-0.50} & -2.50 & \textbf{-0.27} & 0.00   & \textbf{-0.73} & -5.17 & \textbf{-0.86} & -22.50 & \textbf{-0.47} & -5.17 & \textbf{-0.57} & -22.33 \\
8  & 1998 & 5  & 1  & \textbf{-0.61} & 2.50  & \textbf{-0.62} & 2.83   & \textbf{-0.78} & 20.17 & \textbf{-0.95} & 2.50   & \textbf{-0.65} & 12.00 & \textbf{-0.76} & 11.67  \\
9  & 1998 & 5  & 4  & \textbf{-0.71} & 6.67  & \textbf{-0.62} & 5.67   & \textbf{-0.78} & 13.00 & \textbf{-0.74} & 13.33  & \textbf{-0.71} & 27.17 & \textbf{-0.54} & 13.17  \\
10 & 1998 & 8  & 26 & \textbf{-0.70} & 0.00  & \textbf{-0.86} & 0.00   & \textbf{-0.96} & 7.33  & \textbf{-0.87} & 0.83   & \textbf{-0.84} & 14.67 & \textbf{-0.88} & -3.33  \\
11 & 1998 & 9  & 24 & \textbf{-0.82} & 9.83  & \textbf{-0.87} & 0.00   & \textbf{-0.69} & 11.83 & \textbf{-0.94} & -0.33  & \textbf{-0.89} & 29.17 & \textbf{-0.91} & 18.33  \\
12 & 1998 & 10 & 23 & \textbf{-0.22} & 6.50  & \textbf{-0.62} & 2.67   & \textbf{-0.98} & 3.17  & \textbf{-0.99} & 3.00   & \textbf{-0.84} & 25.83 & \textbf{-0.86} & 4.33   \\
13 & 1999 & 1  & 12 & \textbf{0.01}  & 0.00  & \textbf{-0.42} & 0.00   & \textbf{-0.94} & 4.83  & \textbf{-0.88} & -21.00 & \textbf{-0.02} & 20.8 & \textbf{-0.76} & 31.00  \\
14 & 1999 & 2  & 17 & \textbf{-0.45} & 3.33  & \textbf{-0.87} & 0.00   & \textbf{-0.62} & 7.00  & \textbf{-0.88} & -0.83  & \textbf{-0.93} & 17.83 & \textbf{-0.91} & 3.50   \\
15 & 1999 & 6  & 26 & \textbf{-0.33} & 4.33  & \textbf{-0.55} & -11.50 & \textbf{-0.76} & 18.67 & \textbf{-0.76} & -11.00 & \textbf{-0.75} & 21.67 & \textbf{-0.84} & -8.17  \\
16 & 2000 & 3  & 18 & \textbf{-0.70} & 0.00  & \textbf{-0.67} & -0.67  & \textbf{-0.68} & 0.50  & \textbf{-0.87} & -2.83  & \textbf{-0.77} & 0.83  & \textbf{-0.73} & -6.67  \\
17 & 2000 & 4  & 6  & \textbf{-0.41} & 4.67  & \textbf{-0.85} & 0.67   & \textbf{-0.82} & 5.50  & \textbf{-0.63} & 2.50   & \textbf{-0.83} & 22.00 & \textbf{-0.88} & 10.83  \\
18 & 2000 & 5  & 23 & \textbf{-0.44} & 5.50  & \textbf{-0.89} & 3.17   & \textbf{-0.69} & 5.50  & \textbf{-0.82} & 0.83   & \textbf{-0.76} & 11.83 & \textbf{-0.90} & 3.50   \\
19 & 2000 & 6  & 8  & \textbf{-0.45} & 3.67  & \textbf{-0.49} & 16.50  & \textbf{-0.86} & 3.67  & \textbf{-0.93} & 3.83   & \textbf{-0.76} & 3.00  & \textbf{-0.67} & -1.33  \\
20 & 2000 & 6  & 25 & \textbf{-0.47} & 0.33  & \textbf{-0.79} & -1.33  & \textbf{-0.96} & -0.17 & \textbf{-0.92} & -8.67  & \textbf{-0.86} & 13.33 & \textbf{-0.84} & 6.67   \\
21 & 2000 & 9  & 17 & \textbf{-0.59} & 0.00  & \textbf{-0.91} & 1.50   & \textbf{-0.88} & 1.33  & \textbf{-0.95} & 1.00   & \textbf{-0.89} & 19.00 & \textbf{-0.95} & 1.17   \\
22 & 2000 & 12 & 22 & \textbf{-0.86} & 3.33  & \textbf{-0.83} & 2.83   & \textbf{-0.86} & 3.83  & \textbf{-0.71} & 2.83   & \textbf{-0.87} & 1.17  & \textbf{-0.84} & -1.00  \\
23 & 2001 & 3  & 27 & \textbf{-0.24} & 0.83  & \textbf{-0.74} & 0.00   & \textbf{-0.14} & 3.17  & \textbf{-0.95} & 0.50   & \textbf{-0.48} & 3.17  & \textbf{-0.58} & -5.33  \\
24 & 2001 & 3  & 30 & \textbf{-0.09} & 0.00  & \textbf{-0.83} & 0.00   & \textbf{-0.95} & 11.67 & \textbf{-0.72} & 0.67   & \textbf{-0.87} & 46.33 & \textbf{-0.88} & 7.67   \\
25 & 2001 & 3  & 30 & \textbf{-0.09} & 1.17  & \textbf{-0.77} & 4.50   & \textbf{-0.89} & 11.50 & \textbf{-0.69} & 8.17   & \textbf{-0.79} & 38.17 & \textbf{-0.89} & -2.00  \\
26 & 2001 & 4  & 4  & \textbf{-0.81} & 5.17  & \textbf{-0.90} & 2.50   & \textbf{-0.63} & 11.00 & \textbf{-0.90} & 2.67   & \textbf{-0.86} & 6.50  & \textbf{-0.91} & 2.33   \\
27 & 2001 & 4  & 8  & \textbf{-0.79} & 3.00  & \textbf{-0.92} & 3.50   & \textbf{-0.94} & 3.67  & \textbf{-0.95} & 3.50   & \textbf{-0.83} & 10.33 & \textbf{-0.93} & 2.17   \\
28 & 2001 & 4  & 28 & \textbf{-0.22} & 0.00  & \textbf{-0.77} & 3.83   & \textbf{-0.99} & 0 & \textbf{-0.82} & 9.33   & \textbf{-0.72} & 43.8 & \textbf{-0.89} & 16.17  \\
29 & 2001 & 5  & 27 & \textbf{-0.73} & 5.50  & \textbf{-0.88} & 6.17   & \textbf{-0.76} & 5.83  & \textbf{-0.92} & 4.83   & \textbf{-0.79} & 19.67 & \textbf{-0.85} & 7.83   \\
30 & 2001 & 8  & 17 & \textbf{-0.23} & 0.00  & \textbf{-0.61} & 1.50   & \textbf{-0.62} & 2.17  & \textbf{-0.73} & -1.17  & \textbf{-0.87} & 2.33  & \textbf{-0.72} & -2.17  \\
31 & 2001 & 10 & 11 & \textbf{-0.45} & 3.67  & \textbf{-0.73} & 2.83   & \textbf{-0.97} & 3.67  & \textbf{-0.90} & 2.83   & \textbf{-0.75} & 24.83 & \textbf{-0.80} & -3.67  \\
32 & 2001 & 10 & 27 & \textbf{-0.22} & 0.00  & \textbf{-0.81} & -0.33  & \textbf{-0.91} & -2.83 & \textbf{-0.91} & 2.17   & \textbf{-0.74} & 5.00  & \textbf{-0.78} & 5.00   \\
33 & 2001 & 11 & 19 & \textbf{0.31}  & 0.00  & \textbf{-0.92} & 3.00   & \textbf{-0.96} & 3.00  & \textbf{-0.94} & 2.83   & \textbf{-0.69} & 2.83  & \textbf{-0.87} & -12.00 \\
34 & 2001 & 11 & 24 & \textbf{-0.42} & 1.00  & \textbf{-0.77} & 2.00   & \textbf{-0.90} & 12.67 & \textbf{-0.89} & 3.50   & \textbf{-0.74} & 15.67 & \textbf{-0.91} & 1.17   \\
35 & 2002 & 3  & 20 & \textbf{-0.28} & 2.50  & \textbf{-0.73} & 0.00   & \textbf{-0.94} & 15.17 & \textbf{-0.96} & 1.50   & \textbf{-0.88} & 3.00  & \textbf{-0.89} & 7.00   \\
36 & 2002 & 4  & 19 & \textbf{0.20}  & 0.00  & \textbf{-0.47} & 0.83   & \textbf{-0.96} & 0.67  & \textbf{-0.91} & 14.17  & \textbf{-0.76} & 2.00  & \textbf{-0.67} & 5.17   \\
37 & 2002 & 9  & 7  & \textbf{-0.49} & 4.00  & \textbf{-0.55} & 1.50   & \textbf{-0.60} & 11.67 & \textbf{-0.97} & 2.67   & \textbf{-0.53} & -4.50 & \textbf{-0.57} & 10.33  \\
38 & 2002 & 10 & 2  & \textbf{-0.20} & 0.00  & \textbf{-0.41} & 12.17  & \textbf{-0.83} & 0.83  & \textbf{-0.81} & 18.50  & \textbf{-0.83} & 18.83 & \textbf{-0.69} & 22.33  \\
39 & 2002 & 12 & 22 & \textbf{-0.25} & 5.33  & \textbf{0.53}  & 7.00   & \textbf{-0.96} & 7.50  & \textbf{-0.86} & -1.50  & \textbf{-0.50} & 11.83 & \textbf{-0.69} & 8.67   \\
40 & 2003 & 3  & 20 & \textbf{-0.86} & 3.83  & \textbf{-0.70} & 1.83   & \textbf{-0.79} & 18.50 & \textbf{-0.86} & 7.00   & \textbf{-0.78} & 9.67  & \textbf{-0.79} & -11.17 \\
41 & 2003 & 10 & 24 & \textbf{-0.27} & 0.00  & \textbf{-0.32} & 0.00   & \textbf{-0.74} & 0.50  & \textbf{-0.91} & 4.67   & \textbf{-0.81} & 1.00  & \textbf{-0.88} & -13.33 \\
42 & 2003 & 11 & 20 & \textbf{-0.46} & 2.33  & \textbf{-0.55} & 1.83   & \textbf{-0.96} & 0.17  & \textbf{-0.96} & 2.33   & \textbf{-0.70} & 6.83  & \textbf{-0.77} & -4.17  \\
43 & 2004 & 7  & 26 & \textbf{-0.83} & 0.00  & \textbf{-0.81} & 0.50   & \textbf{-0.80} & 0.50  & \textbf{-0.92} & 4.00   & \textbf{-0.92} & 8.83  & \textbf{-0.88} & 9.33   \\
44 & 2004 & 11 & 7  & \textbf{-0.25} & 6.33  & \textbf{-0.71} & 7.50   & \textbf{-0.67} & 8.33  & \textbf{-0.75} & 6.33   & \textbf{-0.72} & 6.4 & \textbf{-0.88} & 6.17   \\
45 & 2004 & 11 & 9  & \textbf{-0.46} & 0.00  & \textbf{-0.78} & 0.00   & \textbf{-0.92} & 0.67  & \textbf{-0.70} & 0.50   & \textbf{-0.72} & 6.33  & \textbf{-0.90} & 16.50  \\
46 & 2005 & 1  & 21 & \textbf{-0.89} & 1.00  & \textbf{-0.90} & 1.17   & \textbf{-0.99} & 1.50  & \textbf{-0.94} & 1.17   & \textbf{-0.91} & 10.67 & \textbf{-0.85} & -0.17  \\
47 & 2005 & 5  & 15 & \textbf{-0.59} & 0.00  & \textbf{-0.83} & 0.67   & \textbf{-0.85} & -1.67 & \textbf{-0.93} & 0.67   & \textbf{-0.80} & 25.50 & \textbf{-0.90} & 22.50  \\
48 & 2005 & 5  & 29 & \textbf{-0.59} & 1.83  & \textbf{-0.25} & 0.67   & \textbf{-0.76} & 1.67  & \textbf{-0.88} & 1.00   & \textbf{-0.85} & 1.33  & \textbf{-0.77} & 15.83  \\
49 & 2005 & 8  & 24 & \textbf{-0.62} & 11.50 & \textbf{-0.87} & 2.50   & \textbf{-0.55} & 11.83 & \textbf{-0.81} & 6.50   & \textbf{-0.77} & 11.83 & \textbf{-0.88} & 4.00   \\
50 & 2005 & 9  & 15 & \textbf{-0.35} & 0.00  & \textbf{-0.93} & -3.00  & \textbf{-0.83} & -3.83 & \textbf{-0.90} & -3.00  & \textbf{-0.83} & 4.00  & \textbf{-0.93} & 2.83  \\
\hline
\hline
\multicolumn{4}{|c|}{\textbf{Mean}}  & \textbf{-0.46} & 2.62  & \textbf{-0.69} & 3.08  & \textbf{-0.81} & 6.53 & \textbf{-0.86} & 6.89  & \textbf{-0.77} & 12.77  & \textbf{-0.81} & 9.77  \\

\hline
\label{table:tab_two}
\end{longtable}









\end{document}